\documentclass[twocolumn,prb,showpacs]{revtex4}
\usepackage{graphicx}
\usepackage{dcolumn}
\usepackage{bm}
\usepackage{amsmath}

\begin{document}

\preprint{}
\title{ Transverse magnetic heat transport on the topological surface}
\author{Takehito Yokoyama$^{1}$ and Shuichi Murakami$^{1,2}$}
\affiliation{$^1$Department of Physics, Tokyo Institute of Technology, Tokyo 152-8551,
Japan \\
$^2$PRESTO, Japan Science and Technology Agency (JST), Kawaguchi, Saitama 332-0012, Japan }
\date{\today}

\begin{abstract}
We investigate Nernst-Ettingshausen and thermal Hall effects on the surface of a topological insulator on which a ferromagnet is attached. We present general expressions of the Peltier and the thermal Hall conductivities, which are reduced to  simple forms at low temperatures.  It is shown that the Peltier  and the thermal Hall conductivities show non-monotonous dependence on temperature. At low temperature, they have linear dependence on temperature. From the behavior of the Peltier conductivity  at low temperature, one can estimate the magnitude of the gap induced by time-reversal symmetry breaking. 
Moreover, we find that the Peltier conductivity can be used to map the Berry phase structure. 
These results open up possibility to control transverse heat transport magnetically.

\end{abstract}

\maketitle

%\affiliation{$^1$Department of Physics, Tokyo Institute of Technology, 2-12-1 Ookayama, Meguro-ku, Tokyo 152-8551, Japan \\

Topological insulator provides us with a new state of matter topologically
distinct from the conventional band insulator~\cite{Hasan}. In particular, edge channels or surface states are described by Dirac fermions and
protected by the band gap in  bulk states. As a result,  backward scattering is forbidden by time-reversal symmetry. From the viewpoint of 
spintronics, it offers a unique opportunity to pursue novel functionality since
spin and momentum are tightly related there.
In fact, several proposals have been made such as the quantized
magneto-electric effect~\cite{Qi,Qi2,Essin,Tse,Nomura}, giant spin rotation~\cite{Yokoyama1},
magnetic properties of the surface state~\cite{Liu}, magneto-transport phenomena~\cite{Yokoyama2,Garate,Yokoyama3,Burkov}, thermoelectric properties\cite{Takahashi}, and superconducting proximity effect including Majorana fermions\cite{Fu,Fu2,Akhmerov,Tanaka,Linder}.

%Since the existence of topological insulators has been confirmed, the next stage in this field is to unveil properties unique to topological insulator. 
Among properties unique to Dirac fermions, Berry phase effect is noticeable. Recently, it has been clarified that  Berry phase plays a pivotal role on anomalous transport in ferromagnets.\cite{Xiao} Now, Berry phase effect is recognized
as one of fundamental effects and is responsible for a wide variety of phenomena.\cite{Xiao2}
In  topological insulator, Berry phase effect can manifest itself in physical quantities by attaching ferromagnet or making thin-film topological insulator \cite{Linder2,Liu2,Lu}. 
Since magnetization of ferromagnet can be controlled by an applied magnetic field, one may expect a remarkable Berry phase effect in magnetic transport on the surface of the topological insulator. 
Moreover, topological insulators discovered so far are also known as good thermoelectric materials. 
Therefore, the surface of topological insulator offers a good arena to study spin caloritronics effect.
% cross-correlation of spin and heat current would have a unique property in the topological insulator, and

In this paper, 
we investigate transverse magnetic heat transport on the surface of the topological insulator on which a ferromagnet is attached. We present general expressions of the  Peltier  and the thermal Hall conductivities, which are reduced to simple forms at low temperatures.  It is clarified that the Peltier and the thermal Hall conductivities show non-monotonous dependence on temperature. At low temperature, they have a linear dependence on temperature. From the behavior of the Peltier conductivity  at low temperature, one can estimate the magnitude of the gap induced by time-reversal symmetry breaking. 
In addition, we find that the Peltier conductivity can be used to map the Berry phase structure. 
These results open up possibility to control transverse thermal currents magnetically.

%%%%%%%%%%%%%%%%%%%%%%%%%%%%%%%%%%%%%%%%%%%%%%%%%%%%%%%%%%%%%%%%%%%%%%%%%%%%%%%%%%%%%
\begin{figure}[tbp]
\begin{center}
\scalebox{0.8}{
\includegraphics[width=8.0cm,clip]{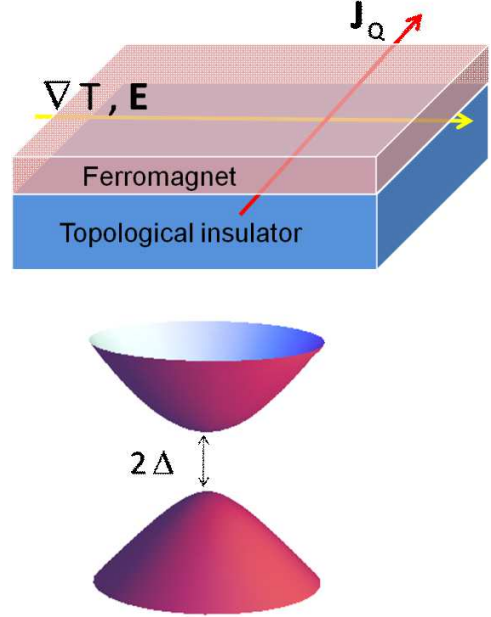}
}
\end{center}
\caption{(Color online) Schematic of the model (upper) and the dispersion of massive Dirac fermion (lower).}
\label{fig1}
\end{figure}

%\begin{figure}[tbp]
%\begin{center}
%\scalebox{0.8}{
%\includegraphics[width=10.5cm,clip]{fig2.eps}}
%\end{center}
%\caption{(Color) Peltier coefficient $\alpha _{xy}$ in unit of $\frac{{k_B e}}{h}$. Thin black lines in the upper panel represent the chemical potential as a function of $T$ for fixed electron densities.  Lower panel: temperature dependence at $\mu/\Delta=2$. The straight line in the lower panel represents the low-temperature asymptotic behavior in  Eq.(\ref{al3}).}
%\label{fig2}
%\end{figure}

%\begin{figure}[tbp]
%\begin{center}
%\scalebox{0.8}{
%\includegraphics[width=10.5cm,clip]{fig3.eps}}
%\end{center}
%\caption{(Color) Normalized thermal Hall conductivity $\kappa _{xy}/\Delta$ in unit of $\frac{{k_B^2}}{h}$. Thin black lines in the upper panel represents the chemical potential as a function of $T$ for fixed electron densities. Lower panel: temperature dependence at $\mu/\Delta=0.5$. The straight line in the lower panel represents the low-temperature asymptotic behavior in  Eq.(\ref{kappa3}).}
%\label{fig3}
%\end{figure}

%%%%%%%%%%%%%%%%%%%%% Formulation
We consider transverse heat transport on the surface of the topological insulator on which the ferromagnet is attached as shown in Fig. \ref{fig1}.  Note that time-reversal symmetry breaking by the ferromagnet is essential to produce Hall effects. 
Heat current under weak electric field $\bm{E}$ and a small thermal gradient $\nabla T$ is determined by 
\begin{equation}
{\bm{J}}_Q  = T\alpha \cdot  {\bm{E}} - \kappa \cdot \nabla T. 
\end{equation}%
Here, we focus on transversal heat transport, namely Nernst-Ettinghausen and Leduc-Righi effects, which are characterized by the Peltier and the thermal conductivity tensors, $\alpha$ and $\kappa$,  respectively. 
These quantities can be calculated as \cite{Bergman,Xiao}
$\alpha _{xy}  = \frac{{k_B e}}{h}c_1 ,\kappa _{xy}  =  - \frac{{k_B^2 T}}{h}c_2 $ where
\begin{eqnarray}
c_i  = \int_{}^{} {\frac{{d{\bm{k}}}}{{(2\pi )^2 }}} \sum\limits_\tau  {\Omega _\tau  } \int_{\varepsilon_{k \tau} - \mu }^\infty  {d\varepsilon (\beta \varepsilon )^i \frac{{\partial f(\varepsilon )}}{{\partial \varepsilon }}}.
\end{eqnarray}
Here, $\Omega _\tau$, $\varepsilon_{k \tau}$, $\mu $ and $f$ are the Berry curvature with the band index $\tau$, the dispersion of the surface Dirac fermion, the chemical potential, and the Fermi distribution function ($f(\varepsilon)=(e^\varepsilon+1)^{-1}$), respectively.
These expressions are reduced to: 
\begin{widetext}
\begin{eqnarray}
\alpha _{xy}  =  - \frac{{k_B e}}{h}\int_{}^{} {\frac{{d{\bm{k}}}}{{(2\pi )^2 }}} \sum\limits_\tau  {\Omega _\tau  } \left[ {\beta (\varepsilon_{k \tau}  - \mu )f(\varepsilon_{k \tau} -\mu) + \log (1 + e^{ - \beta (\varepsilon_{k \tau}  - \mu )} )} \right], \label{al1} \\ 
 \kappa _{xy}  = \frac{{k_B^2 T}}{h}\int_{}^{} {\frac{{d{\bm{k}}}}{{(2\pi )^2 }}} \sum\limits_\tau  {\Omega _\tau  } \left[ {\frac{{\pi ^2 }}{3} + \beta ^2 (\varepsilon_{k \tau}  - \mu )^2 f(\varepsilon_{k \tau}-\mu ) - \left[ {\log (1 + e^{ - \beta (\varepsilon_{k \tau}  - \mu )} )} \right]^2  - 2\rm{Li}_2 (1 - \it{f(\varepsilon_{k \tau} -\mu)})} \right] \label{kappa1} 
\end{eqnarray}
\end{widetext}
where $\rm{Li}$$_2(z)$ is the polylogarithm function.
%\begin{equation}
%\Omega _\tau   =  - \tau \frac{{v_F^2 \Delta }}{{2\left[ {\Delta ^2  + (v_F^{} k)^2 } \right]^{\frac{3}{2}} }}
%\end{equation}%

In the limit of $T \to 0$, we have
\begin{equation}
\alpha _{xy}  =  - \frac{{\pi ^2 }}{3}\frac{{k_B e}}{h}T\int_{}^{} {\frac{{d{\bm{k}}}}{{(2\pi )^2 }}} \sum\limits_\tau  {\Omega _\tau  } \delta (\mu  - \varepsilon_{k \tau} ) \label{al2} 
\end{equation}%
and 
%Also, the thermal Hall conductivity is given by 
\begin{equation}
\kappa _{xy}  = \frac{{\pi ^2 }}{3}\frac{{k_B^2}}{h}T\int_{}^{} {\frac{{d{\bm{k}}}}{{(2\pi )^2 }}} \sum\limits_\tau  {\Omega _\tau  } \theta (\mu  - \varepsilon_{k \tau} ) = \frac{{\pi ^2 }}{3}\frac{{k_B^2}}{e^2}T\sigma _{xy}. \label{kappa2} 
\end{equation}%
Here, $\sigma _{xy}$ is the Hall conductivity. This confirms the Wiedemann-Franz law for the Hall currents. 
We also find the relation
\begin{equation}
\alpha _{xy}   =  - \frac{e}{{k_B^{} }}\frac{{\partial \kappa _{xy} }}{{\partial \mu }}.
\end{equation}%

%The above expression of $\alpha _{xy}$ at low temperature shows that Peltier conductivity can be used to map the Berry phase structure by varying $\mu$. 
The coefficient for the $T$-linear dependence of $\alpha_{xy}$ is the sum of the Berry curvature at the Fermi energy. Therefore, the Peltier conductivity can be used to map the Berry phase structure by varying $\mu$.
On the other hand, the thermal Hall conductivity $\kappa_{xy}$ is proportional to the sum of the Berry curvature over the occupied states, as is similar to the Hall conductivity $\sigma_{xy}$. 
Note that the expressions obtained above are general, applicable to any nearly degenerate band structures, e.g. narrow gap or magnetic semiconductors where the Berry phase plays a significant role.\cite{Nagaosa}

\begin{figure*}[tbp]
\begin{center}
\scalebox{0.8}{
\includegraphics[width=21.0cm,clip]{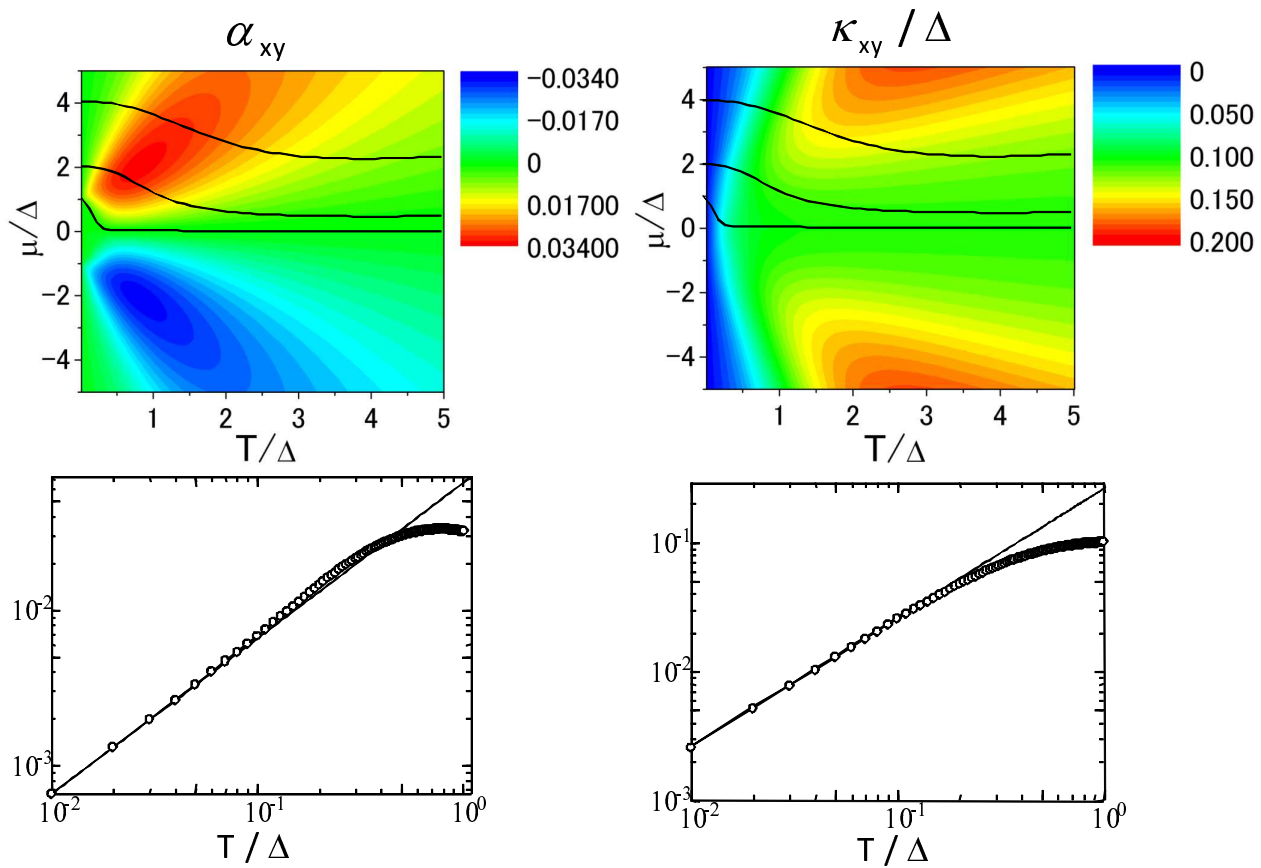}
}
\end{center}
\caption{(Color)  Peltier conductivity $\alpha _{xy}$ in unit of $\frac{{k_B e}}{h}$ (left panels), and normalized thermal Hall conductivity $\kappa _{xy}/\Delta$ in unit of $\frac{{k_B^2}}{h}$ (right panels). Thin black lines in the upper panels represent the chemical potential as a function of $T$ for fixed electron densities. In lower panels, we show  temperature dependence at $\mu/\Delta=2$ (lower left) and that at $\mu/\Delta=0.5$ (lower right). The straight lines in the lower panels represent the low-temperature asymptotic behavior in Eq.(\ref{al3}) and  Eq.(\ref{kappa3}).}
\label{fig4}
\end{figure*}

Now, let us consider the following Hamiltonian as a general model of Dirac fermion on the surface of the topological insulator under the exchange field:
\begin{equation}
H = \sum\limits_{\bm{k}} {c_{\bm{k}}^\dag  \left[ {({\bf{a}} \cdot {\bf{k}})\sigma_x  + ({\bf{b}} \cdot {\bf{k}})\sigma_y  + ({\bf{c}} \cdot {\bf{k}})\sigma_z } \right]c_{\bm{k}}^{} }
\end{equation}%
where $\bf{a}, \bf{b}$ and $\bf{c}$ are real 3D vectors, ${\bf{k}} = \left( {k_x ,k_y ,\Delta } \right)^t $, and $\sigma_{x,y,z}$ are Pauli matrices in spin space. 
Without loss of generality, we can set $c_z  = 1$. 
Now, let us assume two fold rotational symmetry at $k_x=k_y=0$, which imposes $a_z  = b_z  = c_x=c_y=0$. The dispersion is given by $\varepsilon_{k \tau}  = \tau \sqrt {({\bf{a}} \cdot {\bf{k}})^2  + ({\bf{b}} \cdot {\bf{k}})^2  + ({\bf{c}} \cdot {\bf{k}})^2 }$ with $\tau=\pm 1$ which is illustrated in Fig. \ref{fig1}.
After some algebra, we obtain the Berry curvature: 
\begin{equation}
\Omega _\tau   = \frac{{\left( {{\bf{a}} \times {\bf{b}}} \right)_z \Delta }}{{2\varepsilon_{k \tau}^3 }}.
\end{equation}% 
This represents the Berry curvature stemming from Dirac fermion under the exchange field along $z$-axis. Notice that in the presence of the in-plane exchange field, 
%by choosing $z$-axis in $\bf{k}$-space parallel to ${\bf{a}} \times {\bf{b}}$,
by a proper translational operation, 
we can set $a_z  = b_z =0$ in general. 
With a proper transformation, we can derive expressions of the Peltier  and the thermal Hall conductivities by substituting into Eqs.(\ref{al1}) and (\ref{kappa1}) the Berry curvature of the form\cite{Lu,Shan}
\begin{eqnarray}
\Omega _\tau   =  - \tau \frac{\Delta }{{2\left[ {\Delta ^2  + k^2 } \right]^{\frac{3}{2}} }} \label{bp}
\end{eqnarray}%
where $k= \left| \bm{k} \right|$ and $\bm{k}$ represents 2D wavevector: ${\bm{k}} = \left( {k_x , k_y } \right)^t $. 
As is expected, the magnitude of the Berry curvature $\Omega_\tau$ takes its maximum at $k=0$. The extremal values of $\Omega_\tau$ have opposite signs for $\tau=+1$ and $\tau=-1$.

At low temperatures, we obtain a simpler form of $\alpha _{xy}$:
\begin{widetext}
\begin{eqnarray}
\alpha _{xy} = \frac{{\pi ^2 }}{3} \frac{{k_B e}}{h} T\int_{}^{} {\frac{{d{\bm{k}}}}{{(2\pi )^2 }}} \sum\limits_\tau  {\frac{{\tau \Delta }}{{2\left[ {\Delta ^2  + k^2 } \right]^{\frac{3}{2}} }}} \delta (\mu  - \tau \sqrt {\Delta ^2  + k^2 } ) %\nonumber \\ 
 = \tau \frac{\pi }{{12}} \frac{{k_B e}}{h} \frac{\Delta }{{\mu ^2 }}T \label{al3} 
\end{eqnarray}%
when the chemical potential is within the band, and $\alpha _{xy}=0$ otherwise. The thermal Hall conductivity at low temperature is given by\begin{eqnarray}
\kappa _{xy} = - \frac{{\pi ^2 }}{3}\frac{{k_B^2}}{h}T\int_{}^{} {\frac{{d{\bm{k}}}}{{(2\pi )^2 }}} \sum\limits_\tau  {\frac{{\tau \Delta }}{{2\left[ {\Delta ^2  + k^2 } \right]^{\frac{3}{2}} }}} \theta (\mu  - \tau \sqrt {\Delta ^2  + k^2 } ).
\end{eqnarray}%
\end{widetext}
When the chemical potential is inside the gap, the thermal Hall conductivity reads
\begin{eqnarray}
\kappa _{xy}  = \frac{\pi }{{12}}\frac{{k_B^2}}{h}T \label{kappa3}. 
\end{eqnarray}%

In the left panels of Fig. \ref{fig4}, we show the Peltier conductivity $\alpha _{xy}$ numerically calculated with Eq.(\ref{al1}) in unit of $\frac{{k_B e}}{h}$. 
The Peltier conductivity $\alpha_{xy}$ is an odd function of $\mu$.
The sign change across $\mu=0$ reflects the opposite sign of the Berry curvature for the upper and lower bands. We find a region with large magnitude of $\alpha _{xy}$. This can be understood as follows. 
As we noted above, the Peltier conductivity $\alpha_{xy}$ at low $T$ is proportional to the sum of the Berry curvature at the Fermi energy.
At finite temperatures,  thermal broadening changes the $\delta$ function in Eq.(\ref{al3}) into a distribution over the region $\left| {\mu  - \varepsilon_{k \tau} } \right| < aT$ $(a = O(1))$. On the other hand, the Berry curvature takes its maximum at $k=0$. Therefore, when $ - \Delta  < \mu  - aT < \Delta  < \mu  + aT$ is satisfied, the contribution of Berry curvature from the upper band gives large $\alpha_{xy}$. 
On the other hand, if $\mu  - aT <  - \Delta  < \mu  + aT < \Delta$ is satisfied,  the contribution  of Berry curvature  from the lower band dominates, and $\alpha_{xy}$ becomes a large negative value. 
%Large magnitude of  $\alpha _{xy}$ can be expected in these regions. 
This is qualitatively consistent with the upper left panel of Fig. \ref{fig4}.
Thin black lines in the upper panel represent chemical potential as a function of $T$ for fixed electron densities. Along the lines of  fixed electron densities, nonmonotonic dependence on temperature is seen. 
Temperature dependence at $\mu/\Delta=2$ is shown in the lower left panel. 
The straight line in the lower panel represents the low-temperature asymptotic behavior in  Eq.(\ref{al3}). Low temperature behavior is well described by the analytical expression Eq.(\ref{al3}). Thus, from the prefactor of the temperature, one can estimate the magnitude of the gap $\Delta$.

The right panels in Figure \ref{fig4} show normalized thermal Hall conductivity $\kappa _{xy}/\Delta$ in unit of $\frac{{k_B^2}}{h}$, numerically calculated from Eq. (\ref{kappa1}). 
The thermal Hall conductivity $\kappa_{xy}$ is an even function of $\mu$.
The $\kappa _{xy}$ becomes large for large $\mu$ and $\Delta$. 
Due to thermal broadening effect, the step function would be modified in the region $\left| {\mu  - \varepsilon_{k \tau} } \right| < aT$ in Eq.(12).
As we noted earlier, the thermal Hall conductivity $\kappa_{xy}$ is proportional to the sum of the Berry curvature over the occupied states. Since the Berry curvatures for the upper and lower bands have opposite signs, they become small when $\mu$ is well below $-\Delta$ or well above $\Delta$.
For $ - \Delta  < \mu  + aT < \Delta $ or $- \Delta  < \mu  - aT < \Delta $, the contribution  of Berry curvature  from the lower band dominates over that from the upper band. Thus, we expect large thermal Hall conductivity in this region. On the other hand, the prefactor $T$ in Eq.(12) suppresses the magnitude of  $\kappa _{xy}$ at low temperature. In this way, we can understand the behavior of $\kappa _{xy}$ in the upper right panel of Fig. \ref{fig4}. From the plots, we also find $a \sim 3$.
Thin black lines in the upper panel represent chemical potential as a function of $T$ for fixed electron densities. Along the lines of  fixed electron densities, we find non monotonic dependence on temperature. Temperature dependence at $\mu/\Delta=0.5$ is shown in the lower right panel. 
The straight line in the lower panel represents the low-temperature asymptotic behavior in Eq.(\ref{kappa3}), which shows  good agreement with the numerical calculation at low temperature.

%These results open up possibility to control Hall currents magnetically. 
As for experimental realizability, the proximity-induced exchange field is expected to be around 5-50meV. \cite{Chen}
If we set $\Delta=$10 meV, then the parameter range used in this paper is $\mu=-50 \sim 50$ meV and $T=1 \sim 100$K, which are within reach of the present-day technique. 
When thermoelectric effects are measured, there may be additional contributions to the thermoelectric effects such as phonon drag. 
However, these contributions are negligible at low temperatures.

In summary,
we have investigated transverse magnetic heat transport on the surface of topological insulator on which a ferromagnet is attached. We have presented general expressions of the Peltier and the thermal Hall conductivities, which are reduced to simple forms at low temperatures.  It is shown that the Peltier and the thermal Hall conductivities show non-monotonous dependence on temperature. At low temperature, they have a linear dependence on temperature. From the behavior of the Peltier conductivity at low temperature, one can estimate the magnitude of the gap induced by time-reversal symmetry breaking. 
Moreover, we have found that the Peltier conductivity represents the Berry curvature at the Fermi energy and that it can be used to map the Berry phase structure. 
These results open up possibility to control transverse thermal currents magnetically. 

We acknowledge support 
by Grant-in-Aids  
from the Ministry of Education,
Culture, Sports, Science and Technology of Japan
 (No.~21000004 and 22540327).

\end{document}